# A Method for Constructing a Lax Pair for the Ernst Equation

C. J. Papachristou *, B. Kent Harrison **

*Department of Physical Sciences, Naval Academy of Greece, Piraeus 18539, Greece*
*E-mail: papachristou@snd.edu.gr*

**Department of Physics and Astronomy, Brigham Young University, Provo, UT 84602, USA*
*E-mail: bkentharrison@comcast.net, bkh@byu.edu*

**Abstract.** A systematic construction of a Lax pair and an infinite set of conservation laws for the Ernst equation is described. The matrix form of this equation is rewritten as a differential ideal of $gl(2,R)$-valued differential forms, and its symmetry condition is expressed as an exterior equation which is linear in the symmetry characteristic and has the form of a conservation law. By means of a recursive process, an infinite collection of such laws is then obtained, and the conserved "charges" are used to derive a linear exterior equation whose components constitute a Lax pair.

## I. Introduction

The search for the connections between symmetry and integrability has always been a central problem in the study of nonlinear partial differential equations (PDEs). For those PDEs having an underlying variational structure, the work of E. Noether and its extensions (see, e.g., [1, 2]) provide an important link between variational symmetries and conservation laws. Non-variational connections between symmetry and integrability, however, also exist. They are often related to the possibility of "linearizing" a nonlinear PDE by use of a Lax pair, i.e., a pair of coupled PDEs linear in an auxiliary function $\Psi$ and integrable for $\Psi$ on the condition that the original (nonlinear) PDE is satisfied. Linearity is an important issue here, since the symmetry condition (characteristic equation) of a PDE is itself a linear PDE for the symmetry characteristic [1, 2].

A given nonlinear PDE may often be linearized in more than one way by different choices of a Lax pair. A particularly useful choice is the one in which the Lax pair plays the role of a Bäcklund transformation connecting the PDE with its symmetry condition [3], so that the solution $\Psi$ of the pair is a symmetry characteristic for the PDE (or, more generally, is linearly dependent on a symmetry characteristic). Hence, in a sense, the symmetry condition is "built" into the Lax pair. In this way, one obtains a symmetry of the PDE by integrating the associated linear system.

A well-known example where these ideas find wide applications is the self-dual Yang-Mills equation [4, 5]. Interestingly, this has been shown to be a sort of prototype equation from which several other known PDEs are derived by reduction [6, 7]. One such PDE is the Ernst equation of General Relativity describing stationary, axially symmetric gravitational fields. In a previous paper [8] the authors proposed a new Lax pair for this equation (an older one was found by Belinski and Zakharov [9]) and showed that the solution $\Psi$ of this pair is indeed linearly related to a symmetry characteristic. In addition to giving new "hidden" symmetries, the Lax pair also leads to the construction of infinite collections of conservation laws for the Ernst equation.

Admittedly, finding a Lax pair with specific properties almost always requires a certain amount of guessing, as well as a lot of patience in a long trial-and-error process. We now ask the question: Can a linear system such as that of [8] be derived in a systematic way? This article answers this question in the affirmative. As we show, the symmetry condition alone leads one straightforwardly to the discovery of infinite sets of conservation laws, as well as a Lax pair having the desired properties. Our formalism is expressed in the language of exterior differential forms which is both elegant and economical. Hence, for example, differential equations expressing conservation laws, as well as systems of PDEs constituting differential recursion relations or Lax pairs, will now be represented by single exterior equations. In this regard, it would be more appropriate to speak of an *exterior linearization equation*, rather than of a Lax pair in the ordinary sense of this term.

In short, the process is as follows: First, we rewrite the Ernst equation as a differential ideal of matrix-valued differential forms and express its symmetry condition as an exterior equation which is linear in the symmetry characteristic. This latter equation is in conservation-law form, and this fact allows us to introduce a first "conserved charge" or "potential". A second conservation law is then found, with a new potential, and the process continues indefinitely, yielding a double infinity of conserved charges. These charges are related to each other via a certain recursion relation and are used as Laurent coefficients in a series whose terms involve powers (both positive and negative) of a complex "spectral" parameter. This series (assuming it converges) represents some complex function $\Psi$, which is shown to satisfy an exterior linearization equation equivalent to a Lax pair.

## II. Mathematical Preliminaries

The variables $x^\mu \equiv \rho, z$ ($\mu=1, 2$, respectively) will be regarded as local orthogonal coordinates in a 2-dimensional Euclidean space with metric $\delta_{\mu\nu}$. Geometrical objects defined in this space (such as functions or differential forms) are assumed matrix-valued, with values generally in $gl(2,C)$ (with appropriate restrictions, such as real-valuedness, etc., in accordance with physical requirements).

The volume 2-form in our space is

$$\tau = {}^1/_2\, \varepsilon_{\mu\nu}\, dx^\mu dx^\nu = d\rho\, dz$$

(the usual summation convention is assumed). For any 1-form

$$\sigma = \sigma_\mu\, dx^\mu = \sigma_1\, d\rho + \sigma_2\, dz\,,$$

the *dual* of $\sigma$ with respect to $\tau$ is defined as the 1-form $*\sigma$ with components

$$(*\sigma)_\nu = \tau_{\mu\nu}\, \sigma^\mu = \varepsilon_{\mu\nu}\, \delta^{\mu\lambda}\, \sigma_\lambda\,,$$

so that

$$*\sigma = (*\sigma)_\mu\, dx^\mu = -\sigma_2\, d\rho + \sigma_1\, dz\,.$$

In particular, $*d\rho = dz$, $*dz = -d\rho$. Also,

$$*(*\sigma) = -\sigma \tag{2.1}$$

For 1-forms $\sigma_1$ and $\sigma_2$, we have that

$$*\sigma_1 \wedge *\sigma_2 = \sigma_1 \wedge \sigma_2, \quad \sigma_1 \wedge *\sigma_2 = -(*\sigma_1) \wedge \sigma_2 \tag{2.2}$$

We note that the * operation is linear, so that

$$*(\alpha\sigma_1 + \beta\sigma_2) = \alpha *\sigma_1 + \beta *\sigma_2 \tag{2.3}$$

where $\alpha$ and $\beta$ are 0-forms.

Given any differential forms $\zeta$ and $\xi$, we define the commutator

$$[\zeta, \xi] \equiv \zeta \wedge \xi - \xi \wedge \zeta.$$

In particular, if $\sigma$ is a 1-form and $\psi$ is a 0-form, then $[\sigma, \psi] = \sigma\psi - \psi\sigma$ and, by the antiderivation property of the exterior derivative,

$$d[\sigma, \psi] = [d\sigma, \psi] - \{\sigma, d\psi\} \tag{2.4}$$

where, in general, curly brackets denote anticommutators:

$$\{\sigma_1, \sigma_2\} \equiv \sigma_1 \wedge \sigma_2 + \sigma_2 \wedge \sigma_1.$$

We note that, to simplify our notation, we will often omit the symbol $\wedge$ of the exterior product. It should be kept in mind, however, that exterior multiplication of differential forms will always be assumed. Thus, an expression like $\sigma_1 \sigma_2$ should be understood as $\sigma_1 \wedge \sigma_2$.

## III. Ernst Equation: Geometrical Formulation and Symmetry

We adopt the following matrix form of the Ernst equation [6, 7]:

$$(\rho g^{-1} g_\rho)_\rho + (\rho g^{-1} g_z)_z = 0 \tag{3.1}$$

where subscripts denote partial derivatives with respect to the variables $\rho$, $z$, collectively denoted $x^\mu$ ($\mu=1, 2$, respectively). The matrix function $g$ is assumed to be $SL(2,R)$-valued and symmetric. With the parametrization

$$g = \frac{1}{f}\begin{bmatrix} 1 & \omega \\ \omega & f^2 + \omega^2 \end{bmatrix}$$

and by setting $E = f + i\omega$, we recover the Ernst equation in the usual form,

$$(\mathrm{Re}\,E)\nabla^2 E = (\nabla E)^2.$$

With the substitutions

$$A = g^{-1}g_\rho\,,\quad B = g^{-1}g_z\,,$$

equation (3.1) becomes equivalent to the system of PDEs

$$A + \rho\,(A_\rho + B_z) = 0 \tag{3.2a}$$

$$B_\rho - A_z + [A,B] = 0 \tag{3.2b}$$

The second equation is just the integrability condition in order that $g$ may be reconstructed from $A$ and $B$.

We introduce the matrix-valued "connection" 1-form

$$\gamma = g^{-1}dg = A\,d\rho + B\,dz \tag{3.3}$$

The integrability condition $d(dg)=0$ in order that $g$ may be recovered from $\gamma$, together with the obvious requirement that $g$ be nonsingular, yield the Mauer-Cartan equation $\omega=0$, where $\omega$ is the 2-form

$$\omega = d\gamma + \gamma \wedge \gamma = dB\,dz - d\rho\,dA + [A,B]\,d\rho\,dz \tag{3.4}$$

We also construct the 2-form

$$d(\rho\,{*}\gamma) = A\,d\rho\,dz + \rho\,(dA\,dz + d\rho\,dB) \tag{3.5}$$

where ${*}\gamma = -B\,d\rho + A\,dz$.

We now observe that Eqs.(3.2) correspond to the system of exterior equations

$$d(\rho\,{*}\gamma) = 0\,,\quad \omega = 0 \tag{3.6}$$

Indeed, one may consider $d(\rho\,{*}\gamma)$ and $\omega$ as 2-forms in a jet-like space of four variables: the scalar variables $x^\mu=\rho,z$ and the $gl(2,R)$ variables $A$ and $B$. Equations (3.2) are recovered by projecting Eqs.(3.6) onto the base space of the $x^\mu$.

Let $I\{d(\rho\,{*}\gamma), \omega\}$ be the ideal of forms [10-12] generated by the 2-forms $d(\rho\,{*}\gamma)$ and $\omega$. The first form is exact, thus its exterior derivative is trivially a member of the ideal, while, as we can easily show, $d\omega = \omega\wedge\gamma - \gamma\wedge\omega$, which also belongs to $I$. We thus conclude that $I$ is a differential (closed) ideal.

The first of Eqs.(3.6) implies the existence of a matrix potential $X$ such that $\rho\,{*}\gamma=dX$ (that is, $\rho A = X_z$, $\rho B = -X_\rho$). Then, ${*}dX = -\rho\gamma$, and, by the Mauer-Cartan equation $\omega=0$, we get

$$d\rho\,{*}dX - \rho\,d\,{*}dX + dX\,dX = 0 \tag{3.7}$$

[where use has been made of the first of Eqs.(2.2)]. In component form,

$$X_\rho - \rho(X_{\rho\rho} + X_{zz}) + [X_\rho, X_z] = 0 \qquad (3.8)$$

We introduce the covariant derivatives

$$D_\rho = \partial_\rho + [A, \ ], \quad D_z = \partial_z + [B, \ ] \qquad (3.9)$$

(where $\partial_\rho = \partial/\partial\rho$ and $\partial_z = \partial/\partial z$) which are seen to be derivations on the Lie algebra of $gl(2,C)$-valued functions. We also define an exterior covariant derivative $D$ which acts on $gl(2,C)$ functions $\Phi$ as follows:

$$D\Phi = d\Phi + [\gamma, \Phi] = (D_\rho\Phi)d\rho + (D_z\Phi)dz \qquad (3.10)$$

We now look at the symmetry problem for system (3.6). We first note that all symmetries of a system of PDEs can be expressed as infinitesimal transformations of the dependent variables alone [1, 2]. Thus, all symmetries may be represented by "vertical" vector fields, i.e., vectors with vanishing projections on the base space of the $x^\mu$. Let $\delta g = \alpha Q[g]$ be an infinitesimal symmetry transformation of Eq.(3.1), where $\alpha$ is an infinitesimal parameter and $Q$ is a matrix-valued function which may depend locally or nonlocally on $g$. It is convenient to set $Q = g\Phi$, where $\Phi$ is another matrix 0-form. The infinitesimal symmetry of Eq.(3.1) is then written as

$$\delta g = \alpha g \Phi \qquad (3.11)$$

(with appropriate restrictions on $\Phi$ in order that the transformation preserve the symmetric $SL(2,R)$ character of $g$). This induces the symmetry transformations $\delta A = \alpha D_\rho \Phi$, $\delta B = \alpha D_z \Phi$ of system (3.2). These are summarized by the *formal* vector field

$$V = D_\rho \Phi \frac{\partial}{\partial A} + D_z \Phi \frac{\partial}{\partial B} \qquad (3.12)$$

The symmetry condition on the ideal $I$ of the 2-forms $d(\rho *\gamma)$ and $\omega$ is that the Lie derivative with respect to $V$ should leave this ideal invariant [10-12]:

$$L_V I \subset I \ .$$

This is satisfied by requiring that

$$L_V d(\rho *\gamma) = L_V \omega = 0 \quad \mathrm{mod} \quad I\{d(\rho *\gamma), \omega\} \qquad (3.13)$$

By using Eq.(3.4) for $\omega$, taking into account that the Lie derivative commutes with the exterior derivative and satisfies the Leibniz rule, and by noting that

$$L_V \gamma = L_V(Ad\rho + B dz) = (D_\rho\Phi)d\rho + (D_z\Phi)dz = D\Phi = d\Phi + [\gamma, \Phi] \ ,$$

we find that
$$L_V \omega = \omega \Phi - \Phi \omega \equiv [\omega, \Phi],$$

which is automatically a member of the ideal *I*, hence satisfies the condition for *ω* in Eq.(3.13). On the other hand, by noting that

$$L_V *\gamma = L_V(-B\,d\rho + A\,dz) = *D\Phi ,$$

we find that the condition for $d(\rho*\gamma)$ is expressed as an exterior equation which is linear in *Φ*:

$$d(\rho *D\Phi) = 0 \quad \textit{on solutions} \qquad (3.14)$$

(where "on solutions" means: when Eqs.(3.6) are satisfied). In component form,

$$(\rho D_\rho \Phi)_\rho + (\rho D_z \Phi)_z = 0 \quad \textit{on solutions} \qquad (3.15)$$

The reader is invited to derive the symmetry condition (3.15) directly from the Ernst equation (3.1) by assuming a symmetry characteristic $Q=g\Phi$ and by applying the abstract formalism described in [3]. (Note, however, that our present notation is different from that of [3]. Specifically, the symbols $D_\rho$ and $D_z$, which here denote covariant derivatives, have the meaning of *total* derivatives in [3].)

## IV. Conservation Laws and Exterior Linearization Equation

We now turn to integrability characteristics of the Ernst equation. As is well known, the hallmark of integrability is the existence of a linear system or Lax pair. This system may be compactified into a single exterior equation involving 1-forms, which will be referred to as an *exterior linearization equation*. The purpose of this section is to describe a systematic construction of such a linearization equation for the Ernst equation, or equivalently, for the exterior system (3.6).

We begin with the symmetry condition (3.14):

$$d(\rho *D\Phi) = 0 \qquad (4.1)$$

The corresponding infinitesimal symmetry transformation is $g' = g + \alpha g\Phi$, according to Eq.(3.11). This means that $g'$ is a solution of the general PDE (3.1) when $g$ is a solution. However, we will not require here that the new solution $g'$ conform to the extra physical restrictions imposed on the original solution $g$, namely, of being symmetric and having unit determinant. Thus, all real solutions *Φ* of the exterior equation (4.1) will be admissible (e.g., $\Phi = g^{-1}g_z = B$).

As its component form (3.15) suggests, the exterior equation (4.1) expresses a conservation law valid for solutions of the Ernst equation. Equation (4.1) also implies the existence of a "conserved charge" or "potential" *Φ′*, such that

$$d\Phi' = \rho *D\Phi = \rho (*d\Phi + [*\gamma, \Phi])$$

[where use has been made of the linearity property (2.3) of the star operation]. Starring this equation, solving for $d\Phi$, and requiring that $d(d\Phi)=0$, we find another conservation law:

$$d(\rho * D\Phi' - 2\Phi'dz) = 0 ,$$

by which we introduce a new potential $\Phi''$ such that

$$d\Phi'' = \rho * D\Phi' - 2\Phi'dz = \rho(*d\Phi' + [*\gamma, \Phi']) - 2\Phi'dz .$$

Starring this and applying $d(d\Phi')=0$, we obtain yet another conservation law:

$$d(\rho * D\Phi'' - 4\Phi''dz) = 0 , \quad \text{etc.}$$

This process suggests that we consider the following exterior recursion relation:

$$\begin{aligned} d\Phi^{(n+1)} &= \rho * D\Phi^{(n)} - 2n\Phi^{(n)}dz \\ &= \rho(*d\Phi^{(n)} + [*\gamma, \Phi^{(n)}]) - 2n\Phi^{(n)}dz \end{aligned} \quad (4.2)$$

with $\Phi^{(0)}=\Phi$ representing a symmetry characteristic of the Ernst equation in its general form (3.1) [i.e., a solution of Eq.(4.1)].

In order that the exterior equation (4.2) be integrable for $\Phi^{(n+1)}$ for an already known $\Phi^{(n)}$, the integrability condition $d(d\Phi^{(n+1)})=0$ must be satisfied. This yields

$$d\left(\rho * D\Phi^{(n)} - 2n\Phi^{(n)}dz\right) = 0 \quad (4.3)$$

We will now show that Eq.(4.3) is a conservation law valid for solutions of the Ernst equation. The left-hand side of (4.3) is written as

$$\begin{aligned} l.h.s.(4.3) &= d(\rho * d\Phi^{(n)} + [\rho * \gamma, \Phi^{(n)}] - 2n\Phi^{(n)}dz) \\ &= d\rho * d\Phi^{(n)} + \rho d * d\Phi^{(n)} + d[\rho * \gamma, \Phi^{(n)}] - 2n d\Phi^{(n)}dz . \end{aligned}$$

By using property (2.4) and the second property (2.2), we have:

$$\begin{aligned} d[\rho * \gamma, \Phi^{(n)}] &= [d(\rho * \gamma), \Phi^{(n)}] - \rho * \gamma d\Phi^{(n)} - \rho d\Phi^{(n)} * \gamma \\ &= [d(\rho * \gamma), \Phi^{(n)}] + \rho \gamma * d\Phi^{(n)} + \rho * d\Phi^{(n)} \gamma , \end{aligned}$$

$$d\Phi^{(n)}dz = d\Phi^{(n)} * d\rho = d\rho * d\Phi^{(n)} .$$

Therefore,

$$l.h.s.(4.3) = (1-2n)d\rho * d\Phi^{(n)} + \rho d * d\Phi^{(n)} + [d(\rho * \gamma), \Phi^{(n)}] + \rho \gamma * d\Phi^{(n)} + \rho * d\Phi^{(n)} \gamma.$$

Now, by rewriting the recursion relation (4.2) with $(n-1)$ in place of $n$, we can express $d\Phi^{(n)}$, thus also $*d\Phi^{(n)}$, in terms of $\Phi^{(n-1)}$. Substituting for $*d\Phi^{(n)}$ into the expression for the l.h.s. of (4.3), and taking into account that $d\gamma + \gamma\gamma = \omega$, we finally find:

$$l.h.s.(4.3) = [d(\rho * \gamma), \Phi^{(n)}] - \rho^2[\omega, \Phi^{(n-1)}] .$$

We note that this expression vanishes when $d(\rho*\gamma)=0$ and $\omega=0$, i.e., for solutions of the Ernst equation. This proves the conservation-law property of Eq.(4.3).

As we have just shown, the conservation law (4.3) is the necessary condition for $\Phi^{(n)}$ in order that the exterior equation (4.2) be integrable for $\Phi^{(n+1)}$. For $n=0$, Eq.(4.3) is just the symmetry condition (4.1), which is indeed satisfied by $\Phi^{(0)}$ since the latter is, by assumption, a symmetry characteristic. Now, we must show that the solution $\Phi^{(n+1)}$ of Eq.(4.2) also conforms to condition (4.3) with $(n+1)$ in place of $n$. This will ensure that the recursive process may continue indefinitely for all values of $n$, yielding an infinite number of conservation laws from any given symmetry characteristic $\Phi^{(0)}$. This time we need to eliminate $\Phi^{(n)}$ from Eq.(4.2) in favor of $\Phi^{(n+1)}$. By this process we will actually derive the necessary condition for $\Phi^{(n+1)}$ in order that the exterior equation (4.2) be integrable for $\Phi^{(n)}$ when $\Phi^{(n+1)}$ is already known. This will allow us to use the recursion relation (4.2) "backwards" to obtain potentials $\Phi^{(n)}$ and corresponding conservation laws (4.3) for *negative* values of $n$ also. Thus, the validity of Eqs.(4.2) and (4.3) will be extended to all integral values $n=0, \pm 1, \pm 2, \ldots$

Starring Eq.(4.2) and solving for $d\Phi^{(n)}$, we get:

$$d\Phi^{(n)} = -\frac{1}{\rho} *d\Phi^{(n+1)} - [\gamma, \Phi^{(n)}] + \frac{2n}{\rho}\Phi^{(n)} d\rho \qquad (4.4)$$

We apply the integrability condition $d(d\Phi^{(n)})=0$, and use Eq.(4.4) again to replace $d\Phi^{(n)}$ where it appears. Then, a lengthy but relatively straightforward calculation, performed with the aid of properties (2.2) and (2.4), shows that

$$d\left(\rho *D\Phi^{(n+1)} - 2(n+1)\Phi^{(n+1)} dz\right) = [d(\rho*\gamma), \Phi^{(n+1)}] - \rho^2 [\omega, \Phi^{(n)}] \;.$$

So, the left-hand side of the above equation vanishes for solutions of the Ernst equation, as it should.

In conclusion, starting with any symmetry characteristic $\Phi^{(0)}$, we can use the recursion relation (4.2) to find a double infinity of conserved charges (potentials) $\Phi^{(n)}$ for $n= \pm 1, \pm 2, \ldots$ These charges are increasingly nonlocal in $g$, since they involve integrals of increasing order of expressions containing the function $g$.

With these charges in hand, we now introduce a complex variable $\lambda$ (to be identified with a spectral parameter) and construct a function $\Psi(x^\mu, \lambda)$ having the following series representation for $\lambda \neq 0$:

$$\Psi(x^\mu, \lambda) = \sum_{n=-\infty}^{+\infty} \lambda^n \Phi^{(n)}(x^\mu) \qquad (4.5)$$

We assume that the series (4.5) converges to the function $\Psi$ which is single-valued and analytic (as a function of $\lambda$) in some annular region centered at the origin of the $\lambda$-plane. Hence, Eq.(4.5) represents a Laurent expansion of $\Psi$ in this region.

Multiplying the recursion relation (4.2) by $\lambda^n$, summing over all integral values of $n$, and using Eq.(4.5), we find an exterior equation linear in $\Psi$:

$$\rho *D\Psi - 2\lambda \Psi_\lambda dz = \frac{1}{\lambda} d\Psi \qquad (4.6)$$

or explicitly,

$$\rho \, {}^*d\Psi + [\rho \, {}^*\gamma, \Psi] - 2\lambda \Psi_\lambda \, dz = \frac{1}{\lambda} d\Psi \qquad (4.7)$$

Relation (4.6) is an *exterior linearization equation* for the Ernst equation, equivalent to a Lax pair. Specifically, the exterior equation (4.6), linear with respect to $\Psi$, is integrable for $\Psi$ when the exterior equations (3.6) are satisfied.

The proof of this statement is outlined as follows: The integrability condition for solution of Eq.(4.6) is $d(d\Psi)=0$. So, the exterior derivative of the left-hand side of this equation must vanish. By using algebraic manipulations which are by now familiar to the reader (such as, for example, $\{{}^*\gamma, d\Psi\} = -\{\gamma, {}^*d\Psi\}$, $d\Psi_\lambda \, dz = d\rho \, {}^*d\Psi_\lambda$, etc.), the above requirement leads to the following exterior equation:

$$d\rho \, {}^*d\Psi + \rho \, d\, {}^*d\Psi + [d(\rho \, {}^*\gamma), \Psi] + \rho \{\gamma, {}^*d\Psi\} - 2\lambda \, d\rho \, {}^*d\Psi_\lambda = 0 \qquad (4.8)$$

By starring the linear system (4.7), we find an expression for ${}^*d\Psi$:

$$ {}^*d\Psi = -\lambda \rho \, (d\Psi + [\gamma, \Psi]) + 2\lambda^2 \Psi_\lambda \, d\rho \qquad (4.9)$$

Differentiating this with respect to $\lambda$, we have:

$$ {}^*d\Psi_\lambda = -\rho \, (d\Psi + [\gamma, \Psi]) - \lambda \rho \, (d\Psi_\lambda + [\gamma, \Psi_\lambda]) + 4\lambda \Psi_\lambda \, d\rho + 2\lambda^2 \Psi_{\lambda\lambda} \, d\rho $$

$$(4.10)$$

Substituting Eqs.(4.9) and (4.10) into the integrability condition (4.8), we finally get:

$$[d(\rho \, {}^*\gamma) - \lambda \rho^2 \omega, \Psi] = 0,$$

where $\omega = d\gamma + \gamma\gamma$. The above relation is valid independently of $\Psi$ and $\lambda$ if $d(\rho \, {}^*\gamma) = 0$ and $\omega = 0$, i.e., for solutions of the Ernst equation. This proves that the integrability of the exterior equation (4.6) for $\Psi$ is indeed dependent upon the satisfaction of the Ernst equation.

In component form, Eq.(4.6) is written as a pair of linear first-order PDEs for $\Psi$:

$$\rho D_\rho \Psi - 2\lambda \Psi_\lambda = \frac{1}{\lambda} \Psi_z$$

$$\rho D_z \Psi = -\frac{1}{\lambda} \Psi_\rho$$

$$(4.11)$$

The reader is invited to show that the integrability of system (4.11) for $\Psi$ requires that equation (3.1) is satisfied (see also [8]). Thus, (4.11) represents a Lax pair for the Ernst equation. In fact, this pair is equivalent to that found by different means in [8]. What we have shown is that this system may actually be constructed by a remarkably straightforward process, by starting with the symmetry condition of the field equation.

## V. Connection to Other Linear Systems

It can be shown (see [8, 3]) that, by solving the linear system (4.11) for $\Psi$, for a given solution $g$ of the Ernst equation, one simultaneously obtains an infinitesimal "hidden" symmetry of this equation, given by the expression

$$\delta g = \frac{\alpha}{2\pi i} \int_C \frac{d\lambda}{\lambda} \left( g\Psi(x^\mu, \lambda) + \Psi^T(x^\mu, \lambda) g \right) \quad (5.1)$$

where $\alpha$ is an infinitesimal parameter, $C$ is a positively oriented, closed contour around the origin of the $\lambda$-plane, and $\Psi^T$ denotes the transpose of the matrix $\Psi$. (Here, $g$ is assumed to conform to the physical restrictions of being real, symmetric, and of unit determinant. Moreover, $\Psi$ is required to be traceless and to assume real values when $\lambda$ is confined to the real axis. Then, the new solution $g' = g + \delta g$ obeys the same physical restrictions as $g$.) Since solutions of the system (4.11) [or equivalently, the exterior linearization equation (4.6)] are of importance in this regard, any mechanism for producing as many solutions as possible would be useful. We now exhibit a simple transformation which maps solutions of (a form of) the Belinski-Zakharov (B-Z) linear system [9] into solutions of our linearization equation (4.6).

We recall the exterior linearization equation (4.7):

$$\rho\,(*d\Psi + [*\gamma, \Psi]) - 2\lambda\Psi_\lambda\, dz = \frac{1}{\lambda} d\Psi \quad (5.2)$$

where $\Psi$ conforms to the physical conditions mentioned in the previous paragraph; namely, $tr\Psi = 0$ and $\Psi(x^\mu, \lambda^*) = \Psi^*(x^\mu, \lambda)$ (the asterisk here denotes complex conjugation). On the other hand, a variant form of the B-Z linear system, adapted to the particular form of our equations, is the following:

$$\rho\,(*d\Phi + *\gamma\,\Phi) - 2\lambda\Phi_\lambda\, dz = \frac{1}{\lambda} d\Phi \quad (5.3)$$

Let $\Phi(g;\lambda)$ be a non-singular solution of the exterior equation (5.3) for some solution $g$ of the Ernst equation. We assume that $\Phi$ becomes real for real values of $\lambda$. Consider now the function $\Psi(g;\lambda)$ given by

$$\Psi = \Phi T \Phi^{-1} \quad (5.4a)$$

where $T$ is an arbitrary traceless matrix function of the form

$$T = F\left( z - \frac{\lambda \rho^2}{2} + \frac{1}{2\lambda} \right) \quad (5.4b)$$

subject to the condition that $F$ be real-valued for real values of $\lambda$. It may then be proven that $\Psi(g;\lambda)$ is a solution of the linearization equation (5.2).

Although only a subset of the entirety of solutions of Eq.(5.2) can be produced in this fashion, the transformation (5.4) is an effective way of taking advantage of our knowledge regarding the B-Z formulation for the purpose of finding hidden symmetries of the Ernst equation.

Our method for finding a linear system and an infinite number of nonlocal conserved currents for the Ernst equation is closely related to that of Nakamura [13]. In the latter case, the Lax pair does not contain derivative terms with respect to the spectral parameter. Moreover, the infinite set of conservation laws is accompanied by a corresponding infinite set of nonlocal symmetries, which is not the case with our method for the Ernst equation but which *is* the case with regard to another familiar nonlinear system, the self-dual Yang-Mills (SDYM) equation. To achieve these extra characteristics, however, one has to perform an analytic continuation of $g(\rho, z)$ into complex space and introduce more independent variables. In this way the Ernst equation transforms into a reduced form of the SDYM equation, and the mathematical treatments of these two systems become quite similar.

## VI. Summary

In this article we have pursued our study of the relation between symmetry and integrability characteristics of the Ernst equation. Taking advantage of the conservation-law form of the symmetry condition, we have inductively produced a double infinity of nonlocal conserved charges by means of a recursion relation. These charges were then used as Laurent coefficients in an infinite series whose terms involve powers (both positive and negative) of a complex "spectral" parameter. Within its domain of convergence, this series represents a function $\Psi$ which is seen to satisfy a certain linear system, the integrability of which for $\Psi$ is possible in view of the Ernst equation. Finally, we have presented a simple transformation which maps all solutions of the Belinski-Zakharov Lax pair [9] into solutions of our linear system, and we have compared our results to those of Nakamura [13]. Our formalism was developed in the language of differential forms and exterior calculus, which allowed us to present our equations in a more compact, as well as a more elegant form.

It is remarkable that integrability properties of the Ernst equation, such as the existence of Lax pairs and an infinite number of conservation laws, can be derived in a straightforward way by performing rather natural manipulations on the symmetry condition. This characteristic, which is also observed in the case of the SDYM equation, reveals a profound, non-Noetherian connection between symmetry and integrability. It will be further explored in future publications.

## Acknowledgment

We thank Kathleen O'Shea-Arapoglou for kindly reading the manuscript and making several useful suggestions.